\documentclass[useAMS,usenatbib]{mn2e}

\title[]{The Effects of Thermal Conduction on the ADAF with a Toroidal Magnetic Field}
\author[S. Abbassi, J. Ghanbari ,and S. Najar]{S. Abbassi
$^{1}$\thanks{E-mail: sabbassi@dubs.ac.ir}
, J. Ghanbari$^{2}$\thanks{E-mail:ghanbari@ferdowsi.um.ac.ir} and S. Najjar$^3$ \\
$^{1}$Department of Physics, Damghan University of Basic Sciences,
Damghan, Iran\\
$^{2}$Department of Physics, School of Sciences, Ferdowsi
University of Mashhad, Mashhad, 91775-1436, Iran\\
$^{3}$Department of Physics, Islamic Azad University, Mashhad
Branch, Mashhad, Iran \\}
\begin{document}
\date{}

\pagerange{\pageref{firstpage}--\pageref{lastpage}} \pubyear{2004}

\maketitle \label{firstpage}

\begin{abstract}
The observation of the hot gas surrounding Sgr $A^*$ and a few
other nearby galactic nuclei imply that electron and proton mean
free paths are comparable to the gas capture radius. So, the hot
accretion flows is likely to proceed under week-collision
conditions. Hence, thermal conduction has been suggested as a
possible mechanism by which the sufficient extra heating is
provided in hot ADAF accretion disks. We consider the effects of
thermal conduction in the presence of a toroidal magnetic field in
an advection-dominated accretion flow around a compact object. For
a steady-state structure of such accretion flows a set of
self-similar solutions are presented. We find two types solutions
which represent high and slow accretion rate. They have different
behaviors with saturated thermal conduction parameter, $\phi$.
\end{abstract}


\section{INTRODUCTION}
An Advection-Dominated Accretion Flow (ADAF) is defined as one in
which a large fraction of the viscously generated heat is advected
with the accreting gas, and only small fraction of the energy is
radiated. Advection-dominated accretion can occur in two different
limits: 1- At very high mass accretion rates, radiation is trapped
in the accreting gas because of the large optical depth and is
advected with the flow. This limit of advection-dominated,
typically occurs for mass accretion rates $\dot{M}>\dot{M}_{Edd}$
(the Eddington rate). 2- At sufficiently low $\dot{M}$, the
accreting gas can become optically thin. The cooling time of the
gas is then longer than the accreting time, and once again we have
an ADAF.

The thin advection-dominated accretion flow (ADAF) model has been
investigated extensively since the end of 1990s; (Ichimaru 1977,
Narayan $\&$ Yi 1994, 1995a, 1995b; Abramowicz et al. 1995). This
model has been used to interpret the spectra of black hole X-ray
binaries in their quiescent or low/hard state as an alternative to
the Shapiro, Lightman \& Eardly (1976, SLE) solutions. Since ADAFs
have large radial velocities, and infalling matter carries the
thermal energy into the black hole, advective energy transport can
stabilize the thermal instability; thus ADAF models have been
widely applied to explain observations of galactic black hole
candidates (e.g., Narayan et al. 1996; Hameury et al. 1997), the
spectral transition of Cyg X-1 (Esin et al. 1996) and
multi-wavelength spectral properties of Sgr A (Narayan \& Yi
1995b; Manmoto et al. 2000; Narayan et al. 1997). In addition,
many ADAF-like models have been proposed including outflows,
convection, etc.

The theories for the structure and properties of hot accretion
flows has been remained controversial. However, Narayan \& Yi 1994
(hereafter NY1994) derived self-similar ADAF solutions which
emphasized on the importance of the stabilizing role of the radial
heat advection. Subsequent analytical work on the hot accretion
flow has emphasized on the possibility of outflows, motivated by a
positive Bernoulli constant (Narayan \& Yi 1994, Fukue 2004). Also
investigation was indicated a advection-dominated accretion flow,
where the disk is convective in the radial direction (Narayan et
al. 2000, Quatret \& Gruzinov 2000).

The diversity of the models indicate that modelling the hot
accretion flows is a challenging and controversial problem. We
think that, one of the largely neglected physical phenomena is the
thermal conduction while recent observations of the hot accretion
flow around active galactic nuclei indicated that it should be
based on collision-less regime. So, thermal conduction probably
has an important role in energy transport in the accreting
materials in a hot accretion disk where they are nearly completely
ionized. Recently, Medvedev \& Narayan (2001) discovers a new type
of accretion flow, a hot settling flow around a rapidly rotating
neutron star. The flow is cooling-dominated and energetically
similar to the SLE solution. The cooling dominated SLE solution
has been shown to be thermally unstable (Piran 1978, Wandel \&
Liang 1991; Narayan \& Yi 1995a) and, hence, unlikely to exist in
the nature. It has been shown that any accretion flow in which the
heating balances cooling is thermally unstable, if the cooling is
due to the bremsstrahlung emission (Shakura \& Sanyev 1976; Piran
1978). But ADAF is known to be thermally stable (Narayan \& Yi
1995a; Kato et al. 1996, 1997)therefore the cooling is week and
the thermal energy of the flow is not radiated but advected with
the gas. Since the SLE solution for non-ADAF disks is known to be
thermally unstable, one might suspect that the new solution would
also be unstable. However, due to the very high temperature of the
accreting gas, thermal conduction is very strong and could
suppress the thermal instability. So it should be important to
consider the role of thermal conduction in a SLE solution. There
is a branch of solution for the problem of a smooth transition
from an outer Shakura \& Sanyev disk (SSD) to an inner ADAF disk
around a compact object which include an additional mechanism of
energy transport, thermal conduction. A number of authors showed
that the SSD-ADAF transitions were realizable if an extra heat
flux caused by thermal conduction was invoked either in the radial
direction (Honma 1996; Manmoto \& Kato 2000; Gracia et al. 2003)
or in the vertical direction (Meyer \& Meyer-Hofmeister 1994;
Meyer, Liu, \& Meyer-Hofmeister 2000).

The weakly-collisional nature of ADAFs has been noted previously
(Mahadevan \& Quataret 1997). But a few authors tried to study the
role of turbulent heat transport in ADAF disks ( Honma 1996,
Manmoto et al. 2000). Since the thermal conduction acts to oppose
the formation of the temperature gradient that causes it, that
might expect that the temperature and density profile in a thermal
conducting disk should be different from the case in which the
disk is not under the influence of thermal conduction. Recently,
Tanaka \& Menou (2006), studied the effect of saturated thermal
conduction on optically thin ADAFs using an extension of
self-similar solution of NY1994. Their solutions suggest that the
thermal conduction may have an important role in the dynamical
behavior of the hot accretion flows and probably is an important
factor to understand the physics of hot accretion disks.

The aim of this work is to consider the possibility of the thermal
conduction in the presence of toroidal magnetic field, which has
been largely neglected ingredient before, could affect the global
properties of the hot accretion flows substantially. The tangled
magnetic fields in accretion flows would likely reduce the
effective mean free paths of particles. The magnitude of this
reduction, which will depends on the field geometry, is still
unknown. In this manuscript we will investigate the effect of
thermal conduction on the physical structure of ADAF-like
accretion flow around a black hole in the presence of a toroidal
magnetic field.

\section{The Basic Equations}

Let us consider a gaseous disk rotating around a Schwarzschild black
hole of mass $M$. The disk is assumed to be in an
advection-dominated regime. We assume a steady axi-symmetry
accretion flow ($\frac{\partial }{\partial
t}=\frac{\partial}{\partial \phi}=0$) and a geometrically thin disk.
In cylindrical coordinates ($r$, $\phi$, $z$), we vertically
integrate the flow equations. Also, we suppose that all flow
variables are only a function of $r$. We ignored the relativistic
effect and we use Newtonian gravity. The disk is supposed to be
turbulent and possesses an effective turbulent viscous. We adopt
$\alpha$-prescription for viscosity of rotating gas. We have assumed
that the generated energy due to viscosity dissipation and heat
conduction into the volume are balanced by advection cooling. The
magnetic field was considered with toroidal configurations.

We can describe the accretion flows by the fundamental governing
equations which are written by the equation of continuity which is
integrated in the vertical direction is expressed for presence
purpose as

\begin{equation}
\frac{1}{r}\frac{d}{dr}(r\Sigma v_r)=2\dot{\rho}H,
\end{equation}
where $v_r$ is the accretion velocity, $\dot{\rho}$ the mass-loss
rate per unit volume, $H$ the disk half-thickness, and $\Sigma$
the surface density, which is defined as $\Sigma=2\rho H$.

 The equation of motion in the radial direction is:
\begin{equation}
v_r\frac{dv_r}{dr}=\frac{v^2_{\phi}}{r}-\frac{GM}{r^2}-\frac{1}{\Sigma}\frac{d}{dr}
(\Sigma
c^2_s)-\frac{c^2_A}{r}-\frac{1}{2\Sigma}\frac{d}{dr}(\Sigma
c^2_A),
\end{equation}
where $v_{\phi}, c_s, c_A, $ are the rotation velocity of gas
disk, sound speed and Alfv\'en speed, respectively. Sound speed is
defied as $c^2_s=\frac{p_{gas}}{\rho}$, $p_{gas}$ being the gas
pressure and Alfv\'en velocity is defined as
$c^2_A=\frac{B^2_{\phi}}{4\pi\rho}=\frac{2p_{mag}}{\rho}$, where
$p_{mag}$ being the magnetic pressure. The first term in the right
hand side of this equation implies the centrifugal force while the
third term implies the pressure gradient force. The fourth and
fifth terms represent the magnetic force.

By integrating the z-component of the momentum equation along the
z-axis, we have hydrostatic balance in the vertical direction as:
\begin{equation}
\frac{GM}{r^3}H^2=c^2_s[1+\frac{1}{2}(\frac{c_A}{c_s})^2]=(1+\beta)c^2_s,
\end{equation}
where $\beta=\frac{p_{mag}}{p_{gas}}=\frac{1}{2}(\frac{c_A}{c_s})^2$
which indicates the importance of magnetic field pressure compared
to gas pressure. We assume that this value is constant through the
disk, but we will show the dynamical properties of the disk for
different values of $\beta$.

For writing the angular momentum equations we need to choose
proper mechanism for viscosity. We adopt $\alpha$-prescription for
viscosity. The $r\phi$-component of the viscosity stress tensor is
proportional to pressure (Shakura $\&$ Sanyev 1973).
\begin{displaymath}
T_{r\phi}=\eta r \frac{d\Omega}{dr}=\alpha p,
\end{displaymath}
where $\eta=\rho\nu$ is the viscosity, $\nu$ is the kinetic
viscosity and $\alpha$ is a viscous parameter. Here we can see
that the viscosity stress tensor is proportional to the gas
pressure or to the total (gas and magnetic) pressure. So, we can
choose two cases: Case 1: when the pressure is assumed to be the
gas pressure (thermal pressure). Case 2: when the pressure is
assumed to be the magnetic pressure plus the gas pressure. For
simplicity in this investigation we use the case 1, we adopt the
form:
\begin{equation}
\nu=\Omega^{-1}_K \alpha(\frac{p}{\rho})=\alpha C_s H,
\end{equation}
This form is so-called $\alpha$-prescription where $\alpha$ is a
constant less than unity (Shukura $\&$ Sanyev 1973). So, the
angular momentum transfer equation is:
\begin{equation}
r\Sigma v_r\frac{d}{dr}(rv_{\phi})=\frac{d}{dr}(\frac{\alpha
\Sigma c^2_s r^3}{\Omega_k}\frac{d\Omega}{dr}),
\end{equation}
where $\Omega$ is angular velocity and
$\Omega_k=\sqrt{\frac{GM}{r^3}}$ is Keplerian angular velocity.

Now we can write the energy equation considering cooling and
heating processes in an ADAF. The energy equation is:

\begin{equation}
\frac{\Sigma v_r}{\gamma-1}\frac{dc^2_s}{dr}+\frac{\Sigma
c^2_s}{r}\frac{d}{dr}(rv_r)=Q_{vis}-Q_{rad}+Q_{cond},
\end{equation}
where $Q_{vis}$, $Q_{cond}$ and $Q_{rad}$  are the locally
released energies due to viscous dissipation, transported by the
thermal conduction and radiative cooling rate, respectively.

The viscosity generates thermal energy through the differential
rotation of turbulently moving gas, and the viscous heating rate
is expressed by:
\begin{displaymath}
Q_{vis}=rT_{r\phi}\frac{d\Omega}{dr}=\Sigma\frac{\alpha c_s^2
r^2}{\Omega_k}(\frac{d\Omega}{dr})^2
\end{displaymath}

The radiated cooling rate is expressed as:
\begin{displaymath}
Q_{rad}=\frac{8acT^4}{3\kappa\rho H},
\end{displaymath}
where $\kappa$ is the opacity of the rotating gas. We assumed that
the disk to be radiation-pressure-dominated with opacity due to
the electron scattering only. In the right hand side of the energy
equation we have:
\begin{displaymath}
Q_{vis}-Q_{rad}+Q_{cond}=Q_{adv}
\end{displaymath}
where $Q_{adv}$ represents the advective transport of energy and is
defined as the difference between the viscous heating rate,
$Q_{vis}$, and radiative cooling rate, $Q_{rad}$, plus the energy
transported by conduction, $Q_{cond}$. We employ the parameter
$f=1-\frac{Q_{rad}}{Q_{vis}}$ to measure the high degree to energy
are transported by radiation. When $f\sim 1$ radiation can be
negligible so it is in advection dominated state while for small $f$
disk is in radiation dominated case. So we can rearrange the right
hand side of Eq.6 to $fQ_{vis}+Q_{cond}$, where $f\leq1$. In
general, $f$ varies with $r$ and depends on the details of heating
and cooling processes. For simplicity, it is assumed a constant.

The energy transport by thermal conduction, $Q_{cond}$, is:
\begin{equation}
Q_{cond}=-\frac{2H}{r}\frac{d}{dr}(r F_s),
\end{equation}
adopted from the formulation of Cowie $\&$ McKee (1977) and Tanaka
\& Menou (2006). The saturated conduction flux is then written as
$F_s=5\phi \rho c^3_s$, where $\phi$ is the saturated constant
(presumably $\leq 1$), $\rho$ is the gas mass density and $c_s$ is
its sound speed. In order to implement the thermal conductivity as
correctly as possible it is essential to know whether the mean
free paths is less than (or comparable to) the length scale of the
temperature gradient. For electron mean free paths which are
greater than the length scale of temperature gradient, the thermal
conductivity is said to be 'saturated'. Because it is a saturated
flux, it no longer explicitly depends on the magnitude of the
temperature gradient but only does on the direction of this
gradient. The heat will flow outward in a hot accretion flow with
a near-virial temperature profile, hence the positive sign was
adopted for $F_s$.

Finally, since we consider the toroidal component for the global
magnetic field of central stars, the induction equation with field
scape can be written as:
\begin{equation}
\frac{d}{dr}(v_r B_{\phi})=\dot{B_{\phi}},
\end{equation}
where $\dot{B_{\phi}}$ is the field escaping or crating rate due to
the magnetic instability or dynamo effect. We can rewrite this
equation as:
\begin{equation}
v_r\frac{dc^2_A}{dr}+c^2_A\frac{dv_r}{dr}-\frac{c^2_Av_r}{r}=
2c^2_A\frac{\dot{B_{\phi}}}{B_{\phi}}-c^2_A\frac{2\dot{\rho
H}}{\Sigma}
\end{equation}

Now we have a set of MHD equations which describe the dynamical
behavior of ADAF flows. The solution of these equations gives us
the dynamical behavior of the disk, which strongly depends on the
viscosity, magnetic field strength, thermal conduction and
advection rate of energy transport.

\section{Self-similar Solutions for ADAFs with Saturated Conduction}
The self-similar solution can not be able to describe the global
behavior of the accretion flow, because in this method there are not
boundary conditions which have been taken into account. However as
long as we are not interested in the solutions near the boundaries,
such solutions describe correctly the true and useful solutions
asymptotically at intermediate areas.

Narayan \& Yi (1994) simplified 2D axi-symmetric steady state ADAF
problem by assuming that the dynamical variables of the flow have
a power law dependence on $r$. This allowed them to evaluate
directly all radial derivations in the equations. Further, they
eliminated $\theta$ by considering a height-integrated set of
equations. The equation has an analytical self-similar solution
which depends only on three parameters: the viscous parameter
$\alpha$, the ratio of specific heats of the accreting gas
$\gamma$, $f$. In our case, we have a magnetic field in the
structure of the ADAF. So, we have an extra free parameter
$\beta$, which introduces the magnetic field strength in our
self-similar solution.

We assume that each physical quantity can be expressed as a power
law of the radial distance, $r^{\nu}$, where power index $\nu$ is
determined for each physical quantities self-consistently. The
solutions are:
\begin{equation}
v_r=-C_1 \alpha \sqrt{\frac{GM}{r}},
\end{equation}
\begin{equation}
v_{\phi}=C_2  \sqrt{\frac{GM}{r}},
\end{equation}
\begin{equation}
c^2_s=\frac{p}{\rho}=C_3 \frac{GM}{r},
\end{equation}
\begin{equation}
C^2_A=\frac{B^2_{\phi}}{4\pi\rho}=2\beta C_3 \frac{GM}{r},
\end{equation}
\begin{equation}\label{sigma}
\Sigma=\Sigma_0 r^{-\frac{1}{2}},
\end{equation}
\begin{equation}
\dot{\rho}=\dot{\rho}_0 r^{-3},
\end{equation}
\begin{equation}
\dot{B_{\phi}}=\dot{B}_0 r^{-\frac{11}{4}},
\end{equation}
where $C1, C2$ and $C3$ are coefficients that we will determine
later. $\dot{\rho}_0$ and $\dot{B}_0$ are constants, which provide
convenient units with which the equations can be written in the
non-dimensional form.

Using these solutions from the continuity, the momentum, angular
momentum, hydrostatic, energy and induction equations, we can
obtain the following system of dimensionless equations, solved for
$C1, C2, C3, f$ and $\phi$:
\begin{equation}\label{rhodot}
\dot{\rho}=0,
\end{equation}
\begin{equation}
\frac{1}{2}\alpha^2C^2_1+C^2_2-1+\frac{1}{2}[3-\beta]C_3=0,
\end{equation}
\begin{equation}
C_1=\frac{3C_3}{2},
\end{equation}
\begin{equation}\label{Aq}
\frac{H}{r}=\sqrt{(1+\beta)C_3},
\end{equation}
\begin{equation}
C^2_2=\frac{3-\gamma}{\gamma-1}\frac{2}{9f}C_1+\frac{40}{9\alpha
f}\frac{-1}{\sqrt{\frac{3}{2}}}\phi\sqrt{C_1},
\end{equation}
\begin{equation}
\dot{B}_0=\frac{5C_1\alpha GM}{4}\sqrt{4\pi\Sigma_0\frac{\beta
C_3}{\sqrt{(1+\beta)C_3}}},
\end{equation}
This solution tends to the solution presented by Akizuki \& Fukue
(2006) for $\phi=0$. Our solution is compatible with the standard
solution of ADAF, Narayan \& Yi 1994, having no mass
loss(\ref{rhodot}). Actually, radial dependence of surface density,
(\ref{sigma}), is a free parameter. For finding a physical solution,
it should be more than -1. We chose it equal to $\frac{-1}{2}$ to
have a physical solution which is compatible with the standard
solution in the case there is no any wind or mass loss effects. When
it is more than $-\frac{1}{2}$, we have mass loss in the disks.
Fukue (2004) solved the standard disk for the case that radial
dependence of surface density is equal to 2. He showed that the
global behavior of the mass loss disk is quite similar to advective
disks. In additions, we can easily see from (\ref{Aq}) that the disk
thickness becomes large due to magnetic pressure.

In the case of very small $\alpha$, we have $\dot{B}_0=0$ which
means that the creation or escape, so the toroidal component of
magnetic field are balanced each other. In this case, the disk
supports itself with rotation, gas and magnetic pressure. In the
case of a finite $\alpha$, the effect of injection of the toroidal
component of magnetic field, can moderate the dynamical behavior of
the disks.

The above equations describe self-similar behavior of the
optically thick advection dominated accretion disk with saturated
thermal conduction in the presence of toroidal magnetic field. In
a finite $\alpha$, we will solve these equations. After some
algebraic manipulations we can find a fourth order equation for
$C_1$ as follow:
\begin{equation}
D^2C^4_1+2BDC^3_1+(B^2-2D)C^2_1-(A^2+2B)C_1+1=0,
\end{equation}
where
\begin{displaymath}
D=\frac{1}{2}\alpha^2,
\end{displaymath}
\begin{displaymath}
B=[\frac{4}{9f}(\frac{1}{\gamma-1}-\frac{1}{2})-\frac{\beta}{3}+1],
\end{displaymath}
\begin{displaymath}
A=\frac{40}{9\alpha f}\sqrt{\frac{2}{3}}\phi,
\end{displaymath}

This Algebraic equation shows that the variable $C_1$ which
determines the behavior of radial velocity depends only on the
$\alpha$, $\phi$, $\beta$ and $f$. Other flows quantity such as
$C_2$ and $C_3$ can be obtained easily from $C_1$.

\input{epsf}
\begin{figure}
\centerline{{\epsfxsize=8cm\epsffile{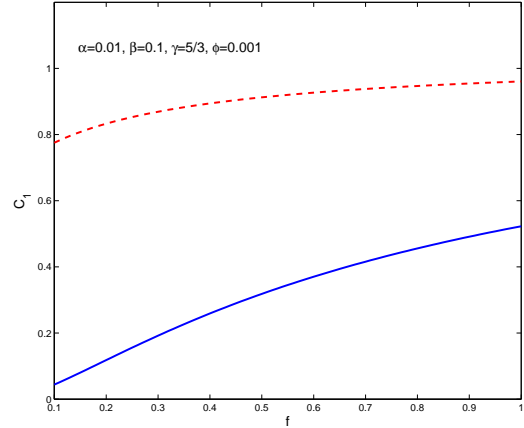}}}
\caption{Profile of $C_1$; coefficient of infall velocity as a
function of $f$ ( \textbf{which represent the importance of
radiation transport}). This profile illustrates the solutions of
forth order equation for $C_1$ in a given range of parameters. The
solid line represents the low accretion rate solution while the
dashed line represents the high accretion rate solutions.}
\end{figure}

This equation is plotted in Fig.1 for the given parameters. We have
found that it has two real roots for this range of parameter space
with two distinct behaviors. One of them represents low accretion
rate and other one represent high accretion rate where both of them
are sub-Keplerian. The parameters of the model are the ratio of
specific heats $\gamma$, the standard viscous parameter $\alpha$,
the radiation transport parameter $f$ and the degree of magnetic
pressure to gas pressure $\beta$.

Figure 1 shows the coefficient $C_1$, which represents the
behavior of the radial flows of accreting materials. Although in
ADAF model the radial infall velocity is generally slower than the
Keplerian speed (=$\sqrt{\frac{GM}{r}}$), it becomes large with
$f$. As the level of advection is increased (by increasing the
$f$), the accreting materials increase their inflow speed in both
two distinct solutions. This variation is around 2 or 3 times for
different range of $f$. This dependence is consistent with the
usual ADAF solutions (Akizuki \& Fukue 2006, Tanaka \& Menou
2006).

\input{epsf}
\begin{figure}
\centerline{{\epsfxsize=9cm\epsfysize=12cm\epsffile{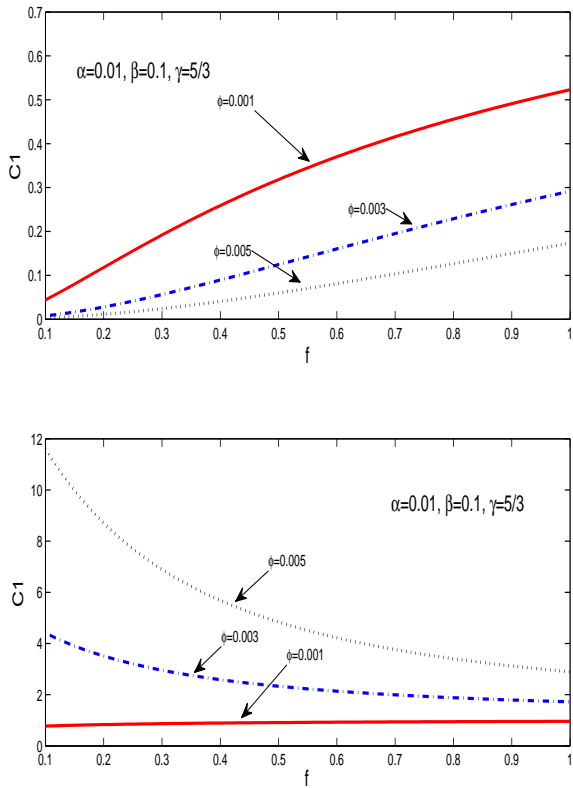}}}
\caption{Numerical coefficient $C_1$ as a function $f$ for
different values of thermal conduction parameter, $\phi$. (Up) for
low accretion rate solution and (Down) for high accretion rate
solution.}
\end{figure}
\input{epsf}

In Fig.2, we investigate the role of saturated thermal conduction on
the radial flow for the two above solutions. These solutions have
different behaviors as we increase the role of thermal conduction by
adding $\phi$. The results for $\phi=0.001$ to $0.006$ are shown in
this manuscript. Tanaka \& Meneu (2006) have shown that for a very
small $\phi$ their solutions coincide the original 1D ADAF
solutions; but by adding saturated conduction parameter, $\phi$, the
effect of thermal conduction can be better seen when we approach to
$\sim 0.001-0.01$. So we have plot our solution in this range. As it
can bee seen in Figure 2 (upper) in the low accretion rate solution,
when we increase $\phi$ the radial flow decreased. But for the high
accretion rate solution the radial flow is increases by increasing
the $\phi$ (Fig.2 lower). Increasing the saturated conduction in
high accretion solution will change the behavior of the radial flow
in different range of advection parameter. The effect of thermal
conduction for low $f$ is larger than compare to $f\sim 1$. The
thermal conduction in low $f$ causes a more radial flow. In the high
accretion solution, by adding the role of advection (adding $f$),
the effect of thermal conduction was reduced. But, for high
accretion solutions, the global effect of thermal conduction
increases the radial flow, whereas in a low accretion solution, this
role is quite suppressed.

To Show the behavior of the solutions, self-similar radial velocity
is demonstrated in Fig.3 for different values $f$ with variation of
viscosity parameters. The most accepted value for $\alpha$ is less
than 0.1. There are probably, more significantly variations when
$\alpha$ exceeds above $0.1$. Recently, King et al. (2007) assert
that in a thin and fully ionized disk, the best observational
evidence suggests a typical range of $\alpha\sim 0.1-0.4$ where
relevant numerical simulations tend to drive the estimates for
$\alpha$ which are one order of magnitude smaller. However, such
large values of $\alpha$ are unlikely (eg. Narayan, Loab, $\&$ Kumar
1994 , Hawley et al. 1994), so we have not explored this region
$\alpha$. Here we investigate these solutions with different values
of viscosity less than 0.1. In the low accretion solution (Fig.3
upper), when we increase the viscosity parameter, we see radial
inflow increases. But in the case of high accretion solution (Fig.3
lower) this behavior is inversed. The behavior of high accretion
solution is similar to the solution presented by Ghanbari et al.
(2007). In this case, Increasing the viscous parameter is
corresponding to the increase of the heating mechanism, so in a
fixed advection regime, there is more energy to advect into the
central star.

Fig.4 shows the coefficient $C_1$ as a function of $f$ for
different values of $\beta$. By adding $\beta$, which indicates
the role of magnetic field in the dynamics of accretion disks, we
will see that the radial flow increases in both solutions. On the
other hand, the radial infall velocity increases when the toroidal
magnetic field becomes large. This is due to the magnetic tension
terms, which dominates the magnetic pressure term in the radial
momentum equation that assists the radial infall motion.

\section{Conclusion}
In this paper we have studied an accretion disk around a black
hole in an advection-dominated regime in the presence of toroidal
magnetic field with thermal conduction. We have presented the
results of self-similar solutions to show the effect of the
viscosity and thermal conduction on magnetically driven accretion
flows from a flow threaded by toroidal magnetic fields. The only
serious approximation we have here made is the use of an isotropic
$\alpha$ viscosity. Attention has restricted to flow accretions in
which self-gravitation is negligible. Considering weekly
collisional nature of hot accretion flows, a saturated form of
thermal conduction has been used as a possible mechanism for
transporting the energy was produced by viscous heating. In a
sense, we have accounted for this possibility by allowing the
saturated conduction constant, $\phi$, to vary in our solutions.

\begin{figure}
\centerline{{\epsfxsize=9cm\epsfysize=12cm\epsffile{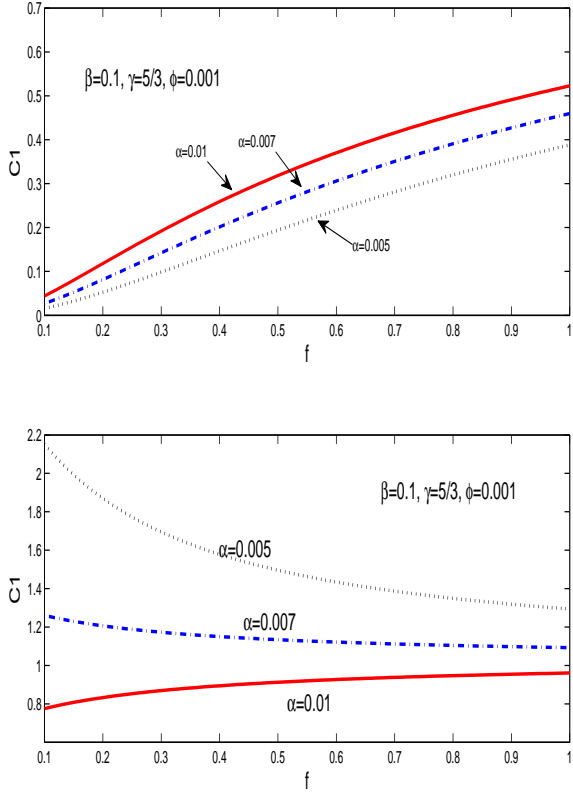}}}
\caption{Numerical coefficient $C_1$ as a function $f$ for
different values of viscous parameter, $\alpha$. (Up) for low
accretion rate solution and (Down) for high accretion rate
solution.}
\end{figure}
\begin{figure}
\centerline{{\epsfxsize=9cm\epsfysize=12cm\epsffile{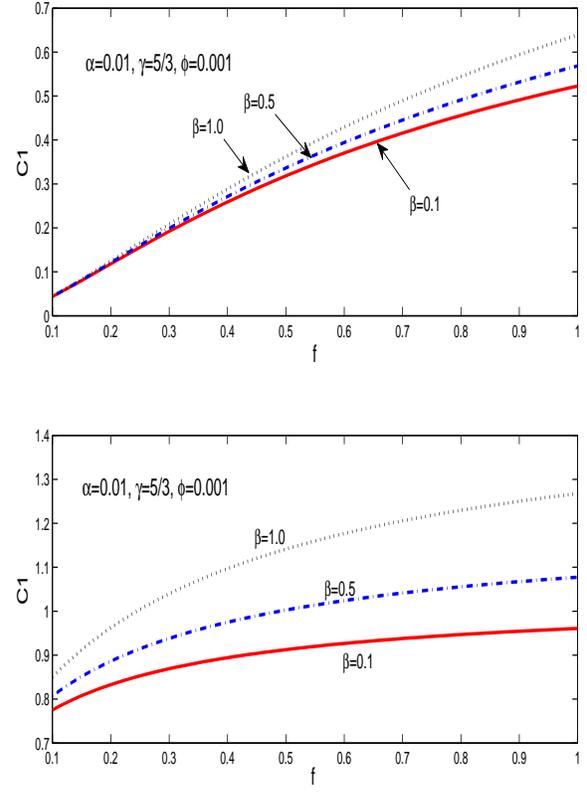}}}
\caption{Numerical coefficient $C_1$ as a function of $f$ for
different values of magnetic field strength, $\beta$. (Up) for low
accretion rate solution and (Down) for high accretion rate
solution.}
\end{figure}

We have found a self-similar solution using similarity technique
in analogy to self-similar solutions by Narayan \& Yi (1994) and
Akizuki \& Fukue (2006). The self-similar solution is valid in the
intermediate region of the advection-dominated disks. The toroidal
geometry for magnetic field is a good approximation for magnetic
structure in this region. The disk structure, is characterized by
the coefficient $C_1$, slightly depends on the strength of
magnetic field. The geometrical thickness of the disks depends the
magnetic field strength; while the structure of accretion flow
will be modified strongly by viscous parameter $\alpha$ and
saturated thermal conduction parameter, $\phi$.

We have found two solutions with different infall velocity, where
they have different behavior with thermal conduction. In the high
accretion rate solution, when the level of saturated conduction
increases, the radial infall velocity also increases, while for
the low accretion rate, the solution is quite different.

The presence of magnetic field with toroidal geometry will affect
the role of thermal conduction. In both solutions, we have found
that a large magnetic field causes a more radial infall velocity.

There are some limitations in our solutions. One of them is that
the self-similar hot accretion flow with conduction is
1-temperature structure. If we use  a 2-temperatures structure for
the ions and electrons in the disks, it is expected that the ions
and electrons temperatures will decouple in the inner regions,
which will modify the role of conduction. The other limitation of
our solution is the anisotropic character of conduction in the
presence of magnetic field. Balbus (2001) has argued that the
dynamical structure of the hot flows could be affected by the
anisotropic character of thermal conduction in the presence of
magnetic field.

However, our results clearly improve the physics of
advection-dominated accretion flow around a black hole. It is
important to investigate the effect of thermal conduction on the
physical structure of hot flow around a black hole. We developed
Narayan \& Yi (1994) standard solutions to a more realistic model
of ADAFs by adding the magnetic field on the structure of disks
and thermal conduction as a mechanism to transport energy in
radial direction. In the future studies, we plan to improve our
model with a more realistic viscous model, Case 2, and to follow
the effect of thermal conduction to other physical parameters of
the disks. Also, we are going to develop our model in a 2-D
self-similar solution for ADAF around a black hole. Several
developments can be investigated to reach a much more realistic
description for the physics of hot accretion disks around a
magnetized compact object.

\end{document}